\documentclass[a4paper,UKenglish]{lipics-v2018}
\usepackage{bussproofs}
\usepackage{amsmath}
\usepackage{microtype}
\usepackage{xspace}
\usepackage{pmboxdraw}
\usepackage{listings}
\usepackage{xcolor}
\usepackage{multicol}
\usepackage{todonotes}
\usepackage{marvosym}
\usepackage{stmaryrd}

\DeclareUnicodeCharacter{2261}{$\equiv$}
\DeclareUnicodeCharacter{27E8}{$\langle$}
\DeclareUnicodeCharacter{27E9}{$\rangle$}
\DeclareUnicodeCharacter{03BC}{$\mu$}
\DeclareUnicodeCharacter{21D2}{$\Rightarrow$}
\DeclareUnicodeCharacter{2200}{$\forall$}
\DeclareUnicodeCharacter{2203}{$\exists$}
\DeclareUnicodeCharacter{2286}{$\subseteq$}
\DeclareUnicodeCharacter{2208}{$\in$}

\lstdefinelanguage{pml}{
  comment=[l]{//},
  morekeywords={def,val,fun,save,restore,let,type,of,rec,case,if,else,fix,
    check,for,because,include,true,false,deduce,show,using,qed,corec,use},
  morestring=[b]",
  literate=
    {∀}{$\forall$}1
    {∃}{$\exists$}1
    {×}{$\times$}1
    {⇒}{$\Rightarrow$}1
    {→}{$\to$}1
    {≡}{$\equiv$}1
    {∈}{$\in$}1
    {μ}{$\mu$}1
    {ν}{$\nu$}1
    {ι}{$\iota$}1
    {ο}{${\rm o}$}1
    {τ}{$\tau$}1
    {κ}{$\kappa$}1
    {⟨}{$\langle$}1
    {⟩}{$\rangle$}1
}

\lstdefinelanguage{none}{
  basicstyle=\small\ttfamily\color{black}
}

\newcommand{\pml}{$\text{PML}_2$\xspace}
\newcommand{\scis}{\LeftScissors}

\title{\pml: Integrated Program Verification in ML}

\author{Rodolphe Lepigre}{LAMA, CNRS, Université Savoie Mont Blanc,
France\\and Inria, LSV, CNRS, Université Paris-Saclay,
France}{rodolphe.lepigre@inria.fr}{https://orcid.org/0000-0002-2849-5338}{}

\authorrunning{R. Lepigre}

\Copyright{Rodolphe Lepigre}

\subjclass{Software and its engineering $\rightarrow$ Software verification}

\EventEditors{Andreas Abel, Fredrik Nordvall Forsberg, and Ambrus Kaposi}
\EventNoEds{3}
\EventLongTitle{23rd International Conference on Types for Proofs and
Programs (TYPES 2017)}
\EventShortTitle{TYPES 2017}
\EventAcronym{TYPES}
\EventYear{2017}
\EventDate{May 29--June 1, 2017}
\EventLocation{Budapest, Hungary}
\EventLogo{}
\SeriesVolume{104}
\ArticleNo{5} 
\nolinenumbers

\begin{document}

\maketitle

\begin{abstract}
  We present the \pml language, which provides a uniform environment for
  programming, and for proving properties of programs in an ML-like setting.
  The language is Curry-style and call-by-value, it provides a control
  operator (interpreted in terms of classical logic), it supports general
  recursion and a very general form of (implicit, non-coercive) subtyping. In
  the system, equational properties of programs are expressed using two new
  type formers, and they are proved by constructing terminating programs.
  Although proofs rely heavily on equational reasoning, equalities are
  exclusively managed by the type-checker. This means that the user only has
  to choose which equality to use, and not where to use it, as is usually done
  in mathematical proofs. In the system, writing proofs mostly amounts to
  applying lemmas (possibly recursive function calls), and to perform case
  analyses (pattern matchings).
\end{abstract}

\section{Introduction: joining programming and proving} 

In the last thirty years, significant progress has been made in the
application of type theory to computer languages. The Curry-Howard
correspondence, which links the type systems of functional programming
languages to mathematical logic, has been explored in two main directions. On
the one hand, proof assistants such as Agda \cite{agda} or Coq \cite{coq} are
based on very expressive logics \cite{mltt, coc}. To establish their
consistency, the underlying programming languages need to be restricted to
provably terminating programs. As a result, they forbid the most general forms
of recursion. On the other hand, functional programming languages such as
Haskell, SML or OCaml are well-suited for programming, as they impose no
restriction on recursion. However, their type systems are inconsistent when
considered as logics, which means that they cannot be used for proving
mathematical formulas.\footnote{This particular point will be explained in
more detail in Section~\ref{termin}.}

The aim of \pml is to provide a uniform environment in which programs can be
designed, specified and proved. The idea is to combine a full-fledged ML-like
programming language, with an enriched type system allowing the specification
of computational behaviours.\footnote{On might argue that \pml is not a
\emph{full-fledged ML-like language} as it does not have mutable references.
It is nonetheless effectful as it provides a control operator similar to
Scheme's \emph{call/cc}.} The obtained system can thus be used as ML for
type-safe general programming, and as a proof assistant for proving properties
of ML programs. The uniformity of the framework implies that programs can be
incrementally refined to obtain more and more guarantees. In particular, there
is no syntactic distinction between programs and proofs. The only difference
is that the latter must be typed-checked against the consistent core of the
system, which only accepts programs that can be proved terminating. In the
current implementation, programs must also be proved terminating to be
directly accepted by the type-checker. It would however be possible to accept
programs that do not pass the termination check, and we would then need to
make sure that such programs are not used to write proofs.\footnote{For
example, we could have two different function types: one used for functions
whose termination has been established, and another one that that may be used
for any function (and that would be a supertype of the former).} Note however
that the system can already be used to reason about arbitrary programs,
including untyped ones and those whose termination cannot be established (an
example will be given in Section~\ref{sect:mc91}).

\subsection{Program verification principles}

In \pml, program properties may be specified with types containing equations
of the form \verb#t ≡ u#, where \verb#t# and \verb#u# are terms of the
language itself. By quantifying over the free variables of these terms, we
can express properties such as the following.
\begin{lstlisting}[]
  // "add" is commutative.
  ∀n∈nat, ∀m∈nat, add n m ≡ add m n

  // "reverse" is involutive.
  ∀a, ∀l∈list⟨a⟩, reverse (reverse l) ≡ l

  // "sort" produces sorted lists.
  ∀l∈list⟨nat⟩, sorted (sort l) ≡ true

  // All natural numbers are equal to "Zero".
  ∀n∈nat, n ≡ Zero.

  // Sorted lists are not affected by "sort".
  ∀l∈list⟨nat⟩, (sorted l ≡ true) ⇒ (sort l ≡ l)
\end{lstlisting}
Of course, such a specification may be inaccurate, in which case it will not
be provable. Note that it is possible to observe complex behaviours using
predicates such as \verb#sorted#, which correspond to boolean-valued
functions.

The \pml language relies on two main ingredients for proving that programs
meet their specifications. First, the type system of the language can be
considered as a (classical) logic through the Curry-Howard correspondence,
although it is only consistent for terminating programs. This (usual)
part of the type system provides basic reasoning principles, which are used
to structure the proofs. The second ingredient is an automatic
decision procedure for the equational theory of the language. It is used to
manage a context of equational assumptions, and eventually prove equations
in this context. The decision procedure is driven by the type-checker,
without any direct interaction with the user. As a consequence, the user only
has to care about the structure of the proof, and not about the details of
where equations should be applied. In fact, equality types of the form
\verb#t ≡ u# are computationally irrelevant in the system. More precisely,
\verb#t ≡ u# is equivalent to the unit type when the denoted equality holds
(and can be proved), and it is empty otherwise. As a consequence, a proof will
generally consist of a (possibly recursive) program that calls other programs
and performs pattern matching but eventually just returns a completely
uninteresting result.  Nonetheless, writing proofs in this way is a very
similar experience to writing functional programs. We can hence hope that our
approach to program verification will feel particularly intuitive to
functional programmers.

\subsection{Previous work on the language}

\pml is based on many ideas introduced by Christophe Raffalli in the PML
language \cite{pml}. Although this first version of the system was very
encouraging on the practical side, it did not stand on solid theoretical
grounds. On the contrary, \pml is based on a call-by-value, classical
realizability model designed by the author \cite{lepigre2016}. This framework
provides a satisfactory way of combining call-by-value evaluation and effects
with dependent function types.\footnote{Due to the soundness issues explained
in previous work \cite{lepigre2016}, the application of dependent functions is
usually restricted to value arguments. This is commonly called \emph{value
restriction} in the context of ML.} The proposed solution relies on a relaxed
form of value restriction (called \emph{semantical value restriction}), which
takes advantage of our notion of program equivalence and its decision
procedure.\footnote{Intuitively, two terms are (observationally) equivalent if
they have the same computational behaviour (i.e., they both converge or they
both diverge) in every possible evaluation context \cite{lepigre2016,
lepigrePhD}.} In particular, it allows the application of a dependent function
to a term which is not a value, under the condition that it can be proved
equivalent to some value. This is especially important because dependent
functions are an essential component of \pml. Indeed, they enable a form of
typed quantification without which many program properties could not be
expressed (see Section~\ref{proofs}).

Another important specificity of \pml's type system is that it relies on the
notion of \emph{local subtyping}, which was introduced in joint work with
Christophe Raffalli \cite{subml}. This framework can be used to give a
syntax-directed formulation of the typing and subtyping rules of the
system\footnote{This means that exactly one typing rule applies for every
term constructor, and only one subtyping rule applies for every pair of type
constructors (up to commutation and circular proof construction).}, despite
its Curry-style nature. In particular, it provides a very general notion of
infinite (circular) proof, that is used for handling inductive and
coinductive types in subtyping derivations, and recursion and termination
checking in typing derivations. Of course, infinite proofs are only valid if
they are well-founded (this is ensured using the size-change principle
\cite{scp}). The combination of local subtyping \cite{subml} and the
realizability model of the system has been addressed in the author's PhD
thesis \cite[Chapter 6]{lepigrePhD}.

Last but not least, the implementation of \pml \cite{implem} was initiated as
part of the author's thesis \cite{lepigrePhD}, and continues with the
collaboration of Christophe Raffalli. The implemented system is intended to
remain very close to the theoretical type system, and every part of the
implementation is justified by the formal semantics of \pml \cite{lepigrePhD}.
Note that all the examples given in this paper are accepted by the version
\verb#2.0.2_types2017# of \pml, which can be downloaded at the following URL.

\begin{center}
  \url{https://github.com/rlepigre/pml/archive/pml_2.0.2_types2017.tar.gz}
\end{center}

\subsection{Disclaimer: the aim of this paper}

This document is intended to be an introductory paper for the \pml system. Its
aim is not to give the details of the realizability semantics, nor to prove
new theoretical results, but rather to list the principles and ideas on
which \pml is based. In particular, Section~\ref{future} contains an
extensive description of several ideas that we would like to investigate in
the near future, and that will be necessary for achieving the goals of \pml
completely. For technical details, the reader should refer to the author's
thesis \cite{lepigrePhD}, and related papers \cite{lepigre2016,subml}.

\section{Functional programming in \texorpdfstring{\pml}{PML2}} 

Our first goal in designing \pml was to obtain a practical, functional
programming language. Out of the many possible technical choices, we decided
to consider a call-by-value language similar to OCaml or SML, as they have
proved to be highly practical and efficient. Our language provides polymorphic
variants \cite{polyvar} and SML-style records, which are convenient for
encoding data types. As an example, the type of lists can be defined as
follows,\footnote{Note that {\verb~list~} is used without its type parameter
in the type of the {\verb~Cns~} constructor. This is due to the fact that the
“{\verb~type rec~}” syntax is desugared to an inductive type (or least fixed
point), and {\verb~list⟨a⟩~} is actually defined as
“{\verb~μ list, [Nil; Cns of \{hd : a; tl : list\}]~}”. In particular, the
current version of \pml does not support polymorphically recursive types.}
together with the corresponding \verb#iter# and \verb#append# functions.
\begin{lstlisting}
type rec list⟨a⟩ = [Nil ; Cns of {hd : a ; tl : list}]

val rec iter : ∀a, (a ⇒ {}) ⇒ list⟨a⟩ ⇒ {} =
  fun f l {
    case l {
      Nil    → {}
      Cns[c] → f c.hd; iter f c.tl
    }
  }

val rec append : ∀a, list⟨a⟩ ⇒ list⟨a⟩ ⇒ list⟨a⟩ =
  fun l1 l2 {
    case l1 {
      Nil    → l2
      Cns[c] → Cns[{hd = c.hd ; tl = append c.tl l2}]
    }
  }
\end{lstlisting}
Note that the \verb#iter# and \verb#append# functions are polymorphic,
which means, for instance, that they can be applied to lists with elements of
an arbitrary type. In our syntax, this is explicitly materialised using
universally quantified type variables. Note also that the \verb#iter#
function relies on the type \verb#{}#, which contains records with no fields.
It plays the same role as OCaml's \verb#unit# type, and its unique inhabitant
is denoted \verb#{}# as well.
\begin{remark}
  As in System F \cite{girard, reynolds}, polymorphism can be used anywhere
  in types, and it is not limited to \emph{let-polymorphism} (or prenex
  polymorphism) as in most ML-like languages.
\end{remark}

\subsection{Control operator and classical logic}\label{intro_classical}

The programming languages of the ML family generally include effectful
operations such as references (i.e., mutable variables). Our system is no
exception since it provides a control operator similar to
\emph{call/cc}.\footnote{This instruction can be used to capture the current
continuation (or evaluation context), so that it can be restored later. It was
first introduced in the Scheme language.} On the programming side, it may be
used to encode a form of exception mechanism. For instance, we can define the
following \verb#exists# function, which tests whether there is an element
satisfying a given predicate in a given list, and stops as soon as possible
if such an element is found.
\pagebreak
\begin{lstlisting}
val exists : ∀a, (a ⇒ bool) ⇒ list⟨a⟩ ⇒ bool =
  fun pred l {
    save k {
      iter (fun e { if pred e { restore k true } }) l;
      false
    }
  }
\end{lstlisting}
Here, the continuation is saved in a variable \verb#k# before calling the
\verb#iter# function, and it is restored with the value \verb#true# if
an element satisfying the predicate is found. In this case, the evaluation
of \verb#iter# is simply aborted. To obtain a similar behaviour without
using a continuation would require the user to write an independent recursive
function (i.e., one that does not rely on \verb#iter#). A more interesting
example that cannot be written without a control operator will be given in
Section~\ref{classical}.

As is now well-known, control operators such as ours can be used to give a
computational content to classical theorems, thus extending the Curry-Howard
correspondence to classical logic \cite{griffin}. It is hence possible to
define programs with a type corresponding to Peirce's law, or to the law of
the excluded middle.
\begin{lstlisting}
val peirce : ∀a b, ((a ⇒ b) ⇒ a) ⇒ a =
  fun x {
    save k {
      x (fun y { restore k y })
    }
  }

// Disjoint sum (logical disjunction) and (logical) negation
type either⟨a,b⟩ = [InL of a ; InR of b]
type neg⟨a⟩ = a ⇒ ∀x,x

val excl_mid : ∀a, {} ⇒ either⟨a, neg⟨a⟩⟩ =
  fun _ {
    save k {
      InR[fun x { restore k InL[x] }]
    }
  }
\end{lstlisting}
Note that the definition of \verb#excl_mid# requires a dummy function
constructor due to the call-by-value evaluation strategy.  Indeed,
\verb#excl_mid# would not be a value if it did not start with an abstraction,
and it would thus save its continuation right away, unlike \verb#peirce#
which must first be given an argument to trigger the computation. This is
related to value restriction \cite{wright1, wright2}, which is required in
presence of control operators \cite{harper}.\footnote{Value restriction is
a sufficient (but not necessary) condition for correctness.}

From a computational point of view, manipulating continuations using
control operators can be understood as ``cheating''. For example
\verb#excl_mid# (or rather, \verb#excl_mid {}#) saves the continuation and
immediately returns a (possibly false) proof of \verb#neg⟨a⟩#. Now, if
this proof is ever applied to a proof of \verb#a# (which would result
in absurdity), the program backtracks and returns the given proof of
\verb#a#. This interpretation has been well-known for a long time, and an
account is given in the work of Wadler \cite[Section 4]{wadler}, for
example.

\subsection{Non-coercive subtyping}

In \pml, subtyping plays a very important role as it allows us to give a
mostly syntax-directed presentation of the system \cite{subml,lepigrePhD}.
Although it is less widespread than polymorphism in mainstream languages,
subtyping can be exploited to improve code modularity. Note that we here
consider a \emph{non-coercive} form of subtyping, which means that if
\verb#a# is a subtype of \verb#b#, then any value of type \verb#a# is also
a value of type \verb#b# (i.e., no coercion is required).

In the system, there are many forms of subtyping that may interact. In
particular, subtyping is used to handle all the connectives that do not
have algorithmic contents (i.e., no counterpart in the syntax of terms).
Such connectives include quantifiers as well as inductive types, but also
the equality types of \pml. Subtyping also plays an important role with
variants and records. For instance, it implies that a record can always
have more fields than required. Moreover, subtyping enables many
commutations of connectives.

As a first example, we can show that the type corresponding to the (classical)
double negation elimination principle can in fact be seen as an instance of
Peirce's law. Indeed, it can be defined as follows in \pml.
\begin{lstlisting}
val dneg_elim : ∀a, neg⟨neg⟨a⟩⟩ ⇒ a =
  peirce
\end{lstlisting}
It is relatively easy to see that the type of \verb#peirce# is indeed a
subtype of that of \verb#dneg_elim#. The corresponding subtyping derivation
is sketched below.\footnote{The proof does not contain all the necessary
information to ensure its validity. The reader should refer to the author's
thesis \cite[Figure 6.5]{lepigrePhD} to fill in the missing details.}
\begin{prooftree}
  \AxiomC{}
  \UnaryInfC{\verb~a0 ⇒ ∀x,x~ \textcolor{red}{$\quad\subseteq\quad$} \verb~a0 ⇒ ∀x,x~}
  \AxiomC{}
  \UnaryInfC{\verb~a0~ \textcolor{red}{$\quad\subseteq\quad$} \verb~a0~}
  \UnaryInfC{\verb~∀x,x~ \textcolor{red}{$\quad\subseteq\quad$} \verb~a0~}
  \BinaryInfC{\verb~(a0 ⇒ ∀x,x) ⇒ ∀x,x~ \textcolor{red}{$\quad\subseteq\quad$}
              \verb~(a0 ⇒ ∀x,x) ⇒ a0~}
  \AxiomC{}
  \UnaryInfC{\verb~a0~ \textcolor{red}{$\quad\subseteq\quad$} \verb~a0~}
  \BinaryInfC{\verb~((a0 ⇒ ∀x,x) ⇒ a0) ⇒ a0~ \textcolor{red}{$\quad\subseteq\quad$}
              \verb~((a0 ⇒ ∀x,x) ⇒ ∀x,x) ⇒ a0~}
  \UnaryInfC{\verb~∀b, ((a0 ⇒ b) ⇒ a0) ⇒ a0~ \textcolor{red}{$\quad\subseteq\quad$}
             \verb~((a0 ⇒ ∀x,x) ⇒ ∀x,x) ⇒ a0~}
  \UnaryInfC{\verb~∀a, ∀b, ((a ⇒ b) ⇒ a) ⇒ a~ \textcolor{red}{$\quad\subseteq\quad$}
             \verb~((a0 ⇒ ∀x,x) ⇒ ∀x,x) ⇒ a0~}
  \UnaryInfC{\verb~∀a, ∀b, ((a ⇒ b) ⇒ a) ⇒ a~ \textcolor{red}{$\quad\subseteq\quad$}
             \verb~∀a, ((a ⇒ ∀x,x) ⇒ ∀x,x) ⇒ a~}
\end{prooftree}
\medskip
Intuitively,
universal quantification on the right of the inclusion can be eliminated by
introducing a fresh constant. On the left, variables that are quantified over
can be replaced by anything. As usual, the subtyping rule for handling the
arrow type reverses the inclusion between the domains due to the
contra-variance of the arrow type.

We will now consider the extension of the type of lists with an additional
constructor allowing constant time concatenation. In \pml, the corresponding
type of ``append lists'' can be defined in such a way that it admits the type
of regular lists as a subtype.
\begin{lstlisting}
type rec alist⟨a⟩ =
  [Nil ; Cns of {hd : a; tl : alist} ; App of alist × alist]

// Constant time "append" function.
val alist_append : ∀a, alist⟨a⟩ ⇒ alist⟨a⟩ ⇒ alist⟨a⟩ =
  fun l1 l2 { App[(l1,l2)] }
\end{lstlisting}
Although regular lists are a special case of ``append lists'', the converse
is not true. To transform an ``append list'' into a list, it is necessary to
define the following recursive, flattening function.
\begin{lstlisting}
val rec alist_to_list : ∀a, alist⟨a⟩ ⇒ list⟨a⟩ = 
  fun l {
    case l {
      Nil    → Nil
      Cns[c] → Cns[{hd = c.hd ; tl = alist_to_list c.tl}]
      App[c] → append (alist_to_list c.1) (alist_to_list c.2)
    }
  }
\end{lstlisting}

Another example of extension for a pre-existing type can be obtained by
defining the type of red-black trees as a subtype of binary trees. More
precisely, a red-black tree can be represented as a tree whose nodes have
an extra color field. Of course, the presence of additional information
in the form of a new record field does not prevent the use of tree
functions such as binary search.

\subsection{Toward the encoding of a module system}

Despite its many different features, \pml remains a fairly small system,
which can be implemented rather concisely. Its design is based on the
principle that every feature should be orthogonal. For instance, there is
only one notion of product type in \pml: records. This is not the case in
OCaml, for instance, which provides tuples, records, objects, modules,
which all have common product type characteristics.

In \pml, modules can be easily encoded using a combination of records for
storing the values, functions for building functors, and existentials for
type abstraction. However, the implementation does not yet provide a
specific syntax for modules. For instance, there is still no way of
``opening'' a module so that its values are accessible in the scope. It
is nonetheless possible to work with the target of the encoding directly.
For example, we can define a type corresponding to the signature (or
interface) of a simple module providing an abstract representation for the
stack data structure with the corresponding operations.\footnote{Here,
$\rm o \to o$ corresponds to the sort of types with one type parameter. In
particular, $\rm o$ is the sort of types (or propositions), and we will
later encounter the sort of program values $\iota$.}
\begin{lstlisting}
type stack_sig = ∃stack: ο → ο,
  { empty : ∀a, stack⟨a⟩;
    push  : ∀a, a ⇒ stack⟨a⟩ ⇒ stack⟨a⟩;
    pop   : ∀a, stack⟨a⟩ ⇒ [None ; Some of a × stack⟨a⟩] }
\end{lstlisting}
An implementation of this interface can then be defined by giving a
corresponding record value. For example, we can implement stacks with
lists as follows.
\begin{lstlisting}
val stack_impl : stack_sig =
  { empty = (Nil : ∀a, list⟨a⟩);
    push  = fun e s { Cns[{hd = e; tl = s}] };
    pop   = fun s {
              case s {
                Nil    → None
                Cns[c] → Some[(c.hd, c.tl)]
              }
            } }
\end{lstlisting}
Note that we need to give at least some type annotation for the system to
know what to instantiate the existential with. This could be done in a
more systematic way with a syntax requiring the user to give the intended
definition for the \verb#stack# type.

\begin{remark}
  It is possible to define a dot-projection operation in order to access
  abstract types, so that it is possible to write \verb#stack_impl.stack#
  to refer to the type of stacks. More details are given in previous work
  \cite{subml}, and in the corresponding implementation.
\end{remark}

\section{Verification of ML programs}\label{proofs} 

\pml is not only a programming language, but also a proof assistant focusing
on program verification. Its proof mechanism relies on equality types of the
form \verb#t ≡ u#, where \verb#t# and \verb#u# are arbitrary (possibly
untyped) terms of the language itself. Such an equality type is inhabited by
the term \verb#{}#\footnote{Recall that it denotes a record with no fields, or
the unique inhabitant of a one-element type.} if the denoted equivalence is
true, and it is empty otherwise.
Equivalences are managed using a partial decision procedure that is driven
by the construction of programs. An equational context is maintained by the
type checker to keep track of the equational assumptions during the
construction of proofs. This context is extended when new equations are
learnt (e.g., when a lemma is applied), and an equation is proved by deriving
a contradiction (e.g., two different variants that are equated) from its
negation.

Terms not only appear in (equality) types, but also play the role of objects
in the underlying logic. In particular, they can be quantified over in types,
and thus form one particular domain of discourse. In fact, our system is based
on a higher-order logic with several atomic sorts (including types and terms),
which means that many different kinds of objects can be quantified over
(universally and existentially) in our types. We can for example
quantify over types with one type parameter (of sort $\rm o \to o$), as in
the signature used for the stack module given in the previous section.

\subsection{(Un)typed quantification and unary natural numbers}

To illustrate the proof mechanism, we will consider very simple examples of
proofs on unary natural numbers. Their type is given below, together with the
corresponding addition function defined using recursion on its first argument.
\begin{lstlisting}
type rec nat = [Zero ; S of nat]

val rec add : nat ⇒ nat ⇒ nat =
  fun n m {
    case n {
      Zero → m
      S[k] → S[add k m]
    }
  }
\end{lstlisting}
As a first example, we will show that for all \verb#n# we have
\verb#add Zero n ≡ n#. This property is expressed using the type
\verb#∀n:#$\iota$\verb#, add Zero n ≡ n#, and it is proved as
follows.\footnote{Here, the domain of the quantification is the
set of values of the language, whose sort is $\iota$. It is not
limited to natural numbers, and also encompasses booleans and
functions for example.}
\begin{lstlisting}
val add_Zero_n : ∀n:ι, add Zero n ≡ n =
  {} // immediate
\end{lstlisting}
The proof is immediate (i.e., only \verb#{}#) as we have \verb#add Zero n ≡ n#
by definition of \verb#add#. Note that this equivalence holds for every value
\verb#n#, whether it corresponds to an element of the type \verb#nat# or not.
For instance, it can be used to show \verb#add Zero true ≡ true# since the
term \verb#add Zero true# evaluates to \verb#true#.
\begin{remark}
  Here, it is crucial that \verb#n# ranges only over values of the language,
  as otherwise the definition of add could not be unfolded. Indeed, since we
  are in call-by-value, it is only possible to effectively apply a function
  when its arguments are all values.
\end{remark}

Let us now show that for every \verb#n# we have \verb#add n Zero ≡ n#.
Although this property looks similar to \verb#add_Zero_n#, the following proof
is invalid.
\begin{lstlisting}
// val add_n_Zero : ∀n:ι, add n Zero ≡ n =
//   {} // invalid
\end{lstlisting}
Indeed, the equivalence \verb#add n Zero ≡ n# does not hold when \verb#n#
is not a unary natural number. In this case, the computation of
\verb#add n Zero# produces a runtime error while that of \verb#n# does
not. As a consequence, we need to
rely on a form of quantification that only ranges over unary natural numbers.
This can be achieved with the type \verb#∀n∈nat, add n Zero ≡ n#, which
corresponds to a (dependent) function taking as input a natural number
\verb#n# and returning a proof of \verb#add n Zero ≡ n#. This property can
then be proved using induction (i.e., using a recursive function) and case
analysis (i.e., pattern matching) with the following program.
\begin{lstlisting}
val rec add_n_Zero : ∀n∈nat, add n Zero ≡ n =
  fun n {
    case n {
      Zero → {}
      S[k] → add_n_Zero k
    }
  }
\end{lstlisting}
If \verb#n# is \verb#Zero#, then we need to show \verb#add Zero Zero ≡ Zero#,
which is immediate by definition of \verb#add#. In the case where \verb#n# is
\verb#S[k]# we need to show \verb#add S[k] Zero ≡ S[k]#. By
definition of \verb#add#, this reduces to \verb#S[add k Zero] ≡ S[k]#.
We can then use the induction hypothesis \verb#add_n_Zero k# to learn
\verb#add k Zero ≡ k# and conclude the proof.
\begin{remark}
  The dependent product type (or typed quantification) constructor
  is not primitive in \pml. It is encoded using a membership type of the form
  \verb#t∈a# which contains all the elements of type \verb#a# that are
  equivalent to the term \verb#t# (it can be seen as a form of singleton
  type). The dependent function type \verb#∀x∈a, b# is then encoded as
  \verb#∀x:#$\iota$\verb#, x∈a ⇒ b#, which corresponds to the relativised
  quantification scheme (see previous work \cite{lepigre2016,lepigrePhD}).
\end{remark}

It is important to note that, in our system, a program that is considered
as a proof needs to go through a termination checker. Indeed, a looping
program could be used to prove anything otherwise.\footnote{More details
will be given in Section~\ref{termin}.} For example, the following
proof is rejected.
\begin{lstlisting}
// val rec add_n_Zero_loop : ∀n∈nat, add n Zero ≡ n =
//   fun n {
//     add_n_Zero_loop n
//   }
\end{lstlisting}
It is however easy to see that \verb#add_Zero_n# and \verb#add_n_Zero# are
terminating, and hence valid. In the following, we will only consider programs
that can be automatically proved terminating by the system.

\subsection{Building up an equational context}

There are two main ways of learning new equations in the system. On the one
hand, when a term \verb#t# is matched in a case analysis, a given branch can
only be reached when the corresponding pattern \verb#C[x]# matches. In this
case we can extend the equational context with \verb#t ≡ C[x]#. On the other
hand, it is possible to invoke a lemma by calling the corresponding function.
In particular, this must be done to use the induction hypothesis in proofs by
induction like in \verb#add_Zero_n# or the following lemma.
\begin{lstlisting}
val rec add_n_S_m : ∀n m∈nat, add n S[m] ≡ S[add n m] =
  fun n m {
    case n {
      Zero → {}
      S[k] → add_n_S_m k m
    }
  }
\end{lstlisting}
In this case, the equation corresponding to the conclusion of the used lemma
is directly added to the context. Of course, more complex results can be
obtained by combining more lemmas. For example, the following proves the
commutativity of addition using a proof by induction with \verb#add_n_Zero#
and \verb#add_n_S_m#.
\begin{lstlisting}
val rec add_comm : ∀n m∈nat, add n m ≡ add m n =
  fun n m {
    case n {
      Zero → add_n_Zero m
      S[k] → add_comm k m; add_n_S_m m k
    }
  }
\end{lstlisting}
\begin{remark}
  Note that terms can be put in sequence with a semicolon. In the above proof,
  the recursive call \verb#add_comm k m# is performed first, before calling
  \verb#add_n_S_m m k#. They are also type-checked in that order, and the
  corresponding equations are added to the context one after the other
  as a side-effect to type-checking. Here, the order in which equations are
  added is not significant (the resulting equational context is the same
  either way), but that is not always the case (lemmas may require some
  equations to hold to be applied).
\end{remark}

\subsection{Detailed proofs using type annotations}

Although the above proof of commutativity is perfectly valid, it might not
be easy enough to read by a human. This problem arises in most proof
assistants. For instance, it is almost impossible to understand a Coq
\cite{coq} proof without replaying it step by step in a compatible editor.
In \pml, it is possible to annotate proofs to highlight the corresponding
thought process. For example, we can reformulate \verb#add_comm# as follows.
\begin{lstlisting}
val rec add_comm : ∀n m∈nat, add n m ≡ add m n =
  fun n m {
    case n {
      Zero → show add Zero m ≡ add m Zero using add_n_Zero m; qed
      S[k] → show add k m ≡ add m k using add_comm k m;
             deduce add S[k] m ≡ S[add m k];
             show add S[k] m ≡ add m S[k] using add_n_S_m m k; qed
    }
  }
\end{lstlisting}
Note that no addition to the system is required for such annotations to be
supported, it is only syntactic sugar. For instance, \verb#qed# is a synonym
of \verb#{}#, and \verb#show u1 ≡ u2 using p# is translated to
\verb#p : u1 ≡ u2#, which amounts to a type coercion.

\begin{remark}
  Many examples of proofs and programs are provided with the implementation
  of the system. Each of the examples given here has been automatically
  checked upon the generation of the document, they are hence correct with
  respect to the implementation.
\end{remark}

\subsection{Mixing proofs and programs}\label{sect:mix}

We will now see that the programming and the proving features of \pml can be
mixed when constructing proofs or programs. In fact, there is no obvious
distinction between the world of the usual programs, and the world of proofs
(remember that proofs are programs in \pml). For instance, it is possible to
combine proofs with programs for them to transport properties (e.g., addition
carrying its own commutativity). This can be achieved using restriction
types, which are in fact used to encode equality types. In \pml, the type
\verb#a | t ≡ u# is equivalent to \verb#a# if \verb#t ≡ u# is true, and to
the empty type otherwise. The type \verb#t ≡ u# is thus encoded as
\verb#{} | t ≡ u#, where \verb#{}# is the unit type. Intuitively, the
restriction type can be seen as a form of conjunction with no algorithmic
contents.

When combined with existential quantification and the membership type,
restriction can be used to encode a \emph{set type} syntax similar to that
of NuPrl \cite{nuprl}. Indeed, we can define \verb#{x ∈ a | t ≡ u}#, which
contains all the elements of type \verb#a# such that \verb#t ≡ u# holds,
as \verb#∃x:#$\iota$\verb#, x∈(a | t ≡ u)#. This provides a very useful
scheme for defining the set of terms of \verb#a# that satisfy some property.
For example, we can encode the type of vectors (i.e., lists of a given length)
by taking every list \verb#l# that has size \verb#s#. The type of vectors
will hence have two parameters: the type of the contained elements and a
term giving the size of vectors.
\begin{lstlisting}
val rec length : ∀a:ο, list⟨a⟩ ⇒ nat =
  fun l {
    case l {
      Nil    → Zero
      Cns[c] → S[length c.tl]
    }
  }

type vec⟨a:ο, s:τ⟩ = {l ∈ list⟨a⟩ | length l ≡ s}
\end{lstlisting}
\begin{remark}
  In the definition of \verb#vec#, the second parameter must have sort $\tau$
  (the sort of terms) and not $\iota$ (the sort of values). Indeed, it is
  often required to work with vectors whose sizes are of the form
  \verb#add n m# (see the definition of the \verb#app# function below).
\end{remark}
\begin{remark}
  There is no constraint on the type of \verb#s# in the definition of
  \verb#vec#. This means that it is possible to consider the type of
  vectors of size \verb#true# for example, but it will be empty since the
  \verb#length# function only returns natural numbers. One of the main
  advantages of this approach is that it is compatible with subtyping.
\end{remark}

Let us stress that vectors can always be used as lists, independently of
their size. The type of vectors is a subtype of the type of lists, as shown
by the following function.
\begin{lstlisting}
val vec_to_list : ∀a:ο, ∀s:τ, vec⟨a,s⟩ ⇒ list⟨a⟩ =
  fun x { x }
\end{lstlisting}
Note that we will never need to use the function \verb#vec_to_list# to
turn a vector into a list. A vector can be seen as a list directly,
without relying on any form of coercion.

We will now define a concatenation function \verb#app# on vectors. It produces
a vector whose length is the sum of the lengths of its two arguments. Note
that we are first required to define the \verb#length_total# function for a
technical reason that will be explained in Section~\ref{termin}.\footnote{We
have good hopes of simplifying this particular point in future work, for
example by automatically obtaining {\verb~length\_total~} from the definition
of {\verb~length~} as they have a similar structure.}
\begin{lstlisting}
val rec length_total : ∀a:ο, ∀l∈list⟨a⟩, ∃v:ι, v ≡ length l =
  fun l {
    case l {
      Nil    → {}
      Cns[c] → length_total c.tl
    }
  }

val rec app : ∀a:ο, ∀m n:ι, vec⟨a, m⟩ ⇒ vec⟨a, n⟩ ⇒ vec⟨a, add m n⟩ =
  fun l1 l2 {
    case l1 {
      Nil    → l2
      Cns[c] → length_total c.tl; Cns[{hd = c.hd; tl = app c.tl l2}]
    }
  }
\end{lstlisting}
Thanks to the Curry-style nature of our system, the sizes of the
argument vectors do not need to be provided as arguments. This may be
surprising for readers that are used to manipulating equivalent types in
Agda or Coq, for example.

\begin{remark}
  In \pml, the proof mechanism can also be used to eliminate unreachable
  code. Indeed, if an equational contradiction is triggered only by
  learning equations along the way, then the code in that branch cannot
  be accessed during evaluation. In this case, a special value \scis{} (to
  be pronounced ``scissors'') can be used. Note that reachability
  information would be particularly useful to efficiently compile \pml
  programs down to assembly code.
\end{remark}

\section{Programs extracted from classical proofs}\label{classical} 

We will now consider an example of a program that can only be written in a
classical setting (i.e., with control operators). We are going to define a
function on streams of natural numbers called \verb#extract#, that extracts
a substream of odd numbers or a substream of even numbers from its input. This
will prove that such a substream exists for all steams of natural
numbers.\footnote{Intuitively, we will have shown that every stream of natural
numbers contains either infinitely many odd numbers or infinitely many even
numbers (and possibly both).}
First, we need to define odd and even numbers using our set type syntax.
\begin{lstlisting}
val rec is_odd : nat ⇒ bool =
  fun n {
    case n {
      Zero → false
      S[m] →
        case m {
          Zero → true
          S[p] → is_odd p
        }
    }
  }

type odd  = {v∈nat | is_odd v ≡ true }
type even = {v∈nat | is_odd v ≡ false}
\end{lstlisting}
As for the \verb#length# function of the previous section, we will need to
show that the \verb#is_odd# function is total for a technical reason (see
Section~\ref{termin} for more details). Intuitively, this will allow us to
reason by cases on the oddness (or evenness) of a given number of the input
stream. Indeed, the totality of \verb#is_odd# implies that this function
always produces a result value, and hence that we can pattern match on its
result.
\begin{lstlisting}
val rec odd_total : ∀n∈nat, ∃v:ι, is_odd n ≡ v =
  fun n {
    case n {
      Zero → {}
      S[m] →
        case m {
          Zero → {}
          S[p] → odd_total p
        }
    }
  }
\end{lstlisting}

We also need to define the type of streams, together with a related type
corresponding to streams with an explicit size annotation (or ordinal)
\verb#s#. Intuitively, this size annotation indicates the number of
elements that are available in the stream (see Section~\ref{termin} for
more details on sized-types).
\begin{lstlisting}
type corec stream⟨a⟩ = {} ⇒ {hd : a; tl : stream}
type sized_stream⟨s,a⟩ = ν_s stream, {} ⇒ {hd : a; tl : stream}
\end{lstlisting}

We can now define the \verb#extract_aux# function, that will be used to
define \verb#extract# on the next page. Note that it relies on \verb#abort#,
which logically amounts to the \emph{ex falso quodlibet} principle. Size
annotations are also required on the type of \verb#extract_aux#, for our
type-checking algorithm to prove its termination.
\begin{lstlisting}
val abort : ∀y, (∀x,x) ⇒ y = fun x { x }

val rec extract_aux : ∀a b,
    neg⟨sized_stream⟨a,even⟩⟩ ⇒
    neg⟨sized_stream⟨b,odd ⟩⟩ ⇒ neg⟨stream⟨nat⟩⟩ =
  fun fe fo s {
    let {hd ; tl} = s {};
    use odd_total hd;
    if is_odd hd {
      fo (fun _ {
        {hd = hd; tl = save oc {
          abort (extract_aux fe (fun x { restore oc x }) tl)}}
      })
    } else {
      fe (fun _ {
        {hd = hd; tl = save ec {
          abort (extract_aux (fun x { restore ec x }) fo tl)}}
      })
    }
  }
\end{lstlisting}
Intuitively, the \verb#extract_aux# function looks at the head of its third
argument (a stream of natural numbers), and depending on whether this number
is odd or even, the function calls one of its first two arguments. They can
be understood as partially constructed stream of even or odd numbers, in the
form of continuations.\footnote{Logical negation is intuitively used to type
continuations represented in the form of a function. A continuation of type
\verb~neg⟨a⟩~ can thus be called with a value of type \verb~a~ in any context
since it yields a logical contradiction (or an element of type \verb~∀x,x~).}
The read number is then added to this stream, and a recursive call is made to
continue the construction.
\begin{remark}
  It may seem surprising that our prototype implementation is able to
  establish the termination of \verb#extract_aux# as an element is added to one of
  two streams at each call. Moreover, this example does not satisfy the
  usually required semi-continuity condition \cite{abel}. It is here accepted
  because our termination test depends more finely on the structure of
  programs than previous approaches \cite{subml}.
\end{remark}

The \verb#extract# function can then be defined as follows, to complete
the construction. The function starts by saving two continuations,
corresponding to the constructors \verb#InL# and \verb#InR# of the return
type, and then calls \verb#extract_aux# on the input stream.
\begin{lstlisting}
val extract : stream⟨nat⟩ ⇒ either⟨stream⟨even⟩, stream⟨odd⟩⟩ =
  fun s {
    save a {
      InL[save ec { restore a InR[save oc {
        abort (extract_aux (fun x { restore ec x})
          (fun x { restore oc x }) s)
      } ] } ]
    }
  }
\end{lstlisting}
The very fact that we can write \verb#extract# proves that it is possible to
extract a stream of odd numbers or a stream of even numbers from any stream of
natural numbers.

Of course, it is only possible to observe a finite prefix of a stream
using a terminating program. As a consequence, we may want to consider
a finite version of \verb#extract#, whose result is a vector of a given
size \verb#n# instead of a stream.
\begin{lstlisting}
val rec prefix : ∀a, ∀n∈nat, stream⟨a⟩ ⇒ vec⟨a,n⟩ =
  fun n s {
    case n {
      Zero → Nil
      S[k] → let {hd ; tl} = s {};
             Cns[{hd ; tl = prefix k tl}]
    }
  }

val finite_extract : ∀n∈nat,
    stream⟨nat⟩ ⇒ either⟨vec⟨even,n⟩, vec⟨odd,n⟩⟩ =
  fun n s {
    case extract s {
      InL[s] → InL[prefix n s]
      InR[s] → InR[prefix n s]
    }
  }
\end{lstlisting}

\begin{remark}
  It is possible to give an equivalent definition of \verb#finite_extract# in an
  intuitionistic setting (i.e., without using a control operator). Indeed,
  at most \verb#2 × n# elements of the input stream need to be considered to
  construct the result.
\end{remark}

\begin{remark}
  Of course, the type of \verb#extract# or \verb#finite_extract# does not directly
  imply that the result of these functions is a substream of their input.
  It is nonetheless easy to convince oneself that this is indeed the case,
  and we could certainly prove it in \pml with some effort.
\end{remark}

To conclude this section, we will consider the result returned by \verb#extract#
(or rather \verb#finite_extract#) on two particular streams. The former will be
the stream of all natural numbers, which can be defined as follows, and is
called \verb#naturals#.
\begin{lstlisting}
val rec naturals_from : nat ⇒ stream⟨nat⟩ =
  fun n _ {
    {hd = n; tl = naturals_from S[n]}
  }

val naturals : stream⟨nat⟩ = naturals_from Zero
\end{lstlisting}
The latter will be a stream of ones, prefixed by three zeroes. It can be
defined as follows, and is called \verb#three_zeroes_then_ones#.
\begin{lstlisting}
val rec ones : stream⟨nat⟩ =
  fun _ { {hd = S[Zero]; tl = ones} }

val three_zeroes_then_ones : stream⟨nat⟩ =
  fun _ { {hd = Zero; tl =
    fun _ { {hd = Zero; tl =
      fun _ { {hd = Zero; tl = ones} }} }} }
\end{lstlisting}
The results of \verb#finite_extract# on \verb#naturals# and
\verb#three_zeroes_then_ones# may be displayed using the following printing
functions.
\begin{lstlisting}
val rec print_nat : nat ⇒ {} =
  fun n {
    case n {
      Zero → print "0"
      S[k] → print "S"; print_nat k
    }
  }

val rec print_list : ∀a, (a ⇒ {}) ⇒ list⟨a⟩ ⇒ {} =
  fun pelt l {
    case l {
      Nil          → print "\n"
      Cns[{hd;tl}] → pelt hd; print " "; print_list pelt tl
    }
  }

val print_res : either⟨list⟨nat⟩, list⟨nat⟩⟩ ⇒ {} =
  fun e {
    case e {
      InL[l] → print "  InL "; print_list print_nat l
      InR[l] → print "  InR "; print_list print_nat l
    }
  }
\end{lstlisting}
\begin{remark}
  Although the above is boiler-plate code, it is provided so that the
  examples in this document are completely self-contained, and can be
  type-checked and evaluated by \pml without any modification.
\end{remark}
We have now all the components that are required to run some tests, and
to display prefixes of the streams produced by the \verb#extract# function.
We will display, for each example, prefixes of increasing size. We will
thus rely on the following \verb#test# function, taking as input a stream
of natural numbers, and showing the result of applying the \verb#finite_extract#
function on this stream with various prefix lengths (from zero to four).
\begin{lstlisting}
val test : stream⟨nat⟩ ⇒ {} =
  fun s {
    print_res (finite_extract Zero s);
    print_res (finite_extract S[Zero] s);
    print_res (finite_extract S[S[Zero]] s);
    print_res (finite_extract S[S[S[Zero]]] s);
    print_res (finite_extract S[S[S[S[Zero]]]] s)
  }
\end{lstlisting}
Let us first consider the output that is produced by applying the above
\verb#test# function to \verb#three_zeroes_then_ones#, which contains
three zeroes followed by infinitely many ones.
\begin{lstlisting}[language=none]
InL 
InL 0 
InL 0 0 
InR S0 S0 S0 
InR S0 S0 S0 S0 
\end{lstlisting}
As one should expect, the computation of the smallest prefixes yields a
list of even numbers. However, if more elements of the input stream are
read, the \verb#extract# function eventually backtracks and produces a list
of odd numbers instead. Indeed, the input stream only contains three
even numbers.

Note that one could expect the fourth line of the output to show a list of
even numbers with three zeroes. The produced result is due to the definition
of \verb#extract#, which looks ahead one element further than strictly necessary
in the input stream. It would be possible to avoid doing so, but the function
would be even more complex than it already is. Note that this also has
consequences on the result obtained by running \verb#test# on \verb#naturals#.
\begin{lstlisting}[language=none]
InL 
InL 0 
InL 0 SS0 
InL 0 SS0 SSSS0 
InL 0 SS0 SSSS0 SSSSSS0 
\end{lstlisting}
Here, one would expect the prefixes to alternate between lists of
even numbers, and lists of odd numbers. Indeed, the stream of all the
natural numbers both contain an infinite sub-stream of even numbers, and
an infinite sub-stream of odd numbers.

\section{Termination and internal totality proofs}\label{termin} 

We will now look more deeply into the relation between proofs and termination
checking in \pml. Technically, the termination of \pml programs (and thus of
\pml proofs) is established using circular proof techniques introduced in
joint work with Christophe Raffalli \cite{subml} and adapted in the author's
thesis \cite{lepigrePhD}. The idea is to type recursive
programs, or more precisely the fixed-point combinator used by \pml, using a
simple unfolding rule. In other words, instances of the fixed-point construction
of the language are typed assuming that they can be typed, thus leading to
a circular structure. Of course, proofs constructed in this way may be
invalid (i.e., not well-founded). To rule out such invalid proofs, a test
based on the size-change principle \cite{scp} is used. When it is able to
show that the structure of a proof is indeed well-founded, termination then
follows from a standard semantic proof by realizability.\footnote{In this
case, adequacy can still be proved by well-founded induction on the structure
of the typing proof.}

\subsection{Termination and consistency}

As mentioned in the introduction, practical functional programming languages
like OCaml or Haskell cannot be used to prove mathematical formulas, since
their type system is not consistent when seen as a logic.
More precisely, the “empty type” is inhabited by a
simple looping program in these systems. Any formula can thus be proved
through the \emph{ex falso quodlibet} principle, as demonstrated by the
following piece of Haskell code.\footnote{Note that the {\verb~Rank2Types~}
(or {\verb~RankNTypes~}) extension is required for the definition of
{\verb~Empty~}.}
\begin{lstlisting}[language=haskell,morekeywords={forall}]
type Empty = forall a. a

bad :: Empty
bad = bad

ex_falso :: Empty → a
ex_falso e = e
\end{lstlisting}
A similar example can also be given in OCaml, but a slightly more complex
definition is required for \verb#bad#.\footnote{Note that {\verb~empty~} is
encoded using a polymorphic record field.} This is due to the call-by-value
evaluation strategy of the language, which restricts the use of the
\verb#let rec# construct to the definition of functions.
\begin{lstlisting}[language=caml]
type empty = { any : 'a.'a }

let bad : empty =
  let rec bad_aux : unit → empty = fun () → bad_aux () in
  bad_aux ()

let ex_falso : type a. empty → a =
  fun e → e.any
\end{lstlisting}
Of course, a similar example can be written in \pml. This is why the
current implementation requires every program (not only proofs) to
pass the termination check.

As \pml can be used to prove program equivalences, the inconsistency that
would be introduced by possible non-termination would allow proving any
program equivalence by inhabiting the corresponding type. Moreover, a
non-terminating program would allow invalid program equivalences to be
added to the equational context. The following invalid program (first given
in Section~\ref{proofs}) gives an example of such a scenario.
\begin{lstlisting}
// val rec add_n_Zero_loop : ∀n∈nat, add n Zero ≡ n =
//   fun n {
//     add_n_Zero_loop n
//   }
\end{lstlisting}
Here, the recursive call \verb#add_n_Zero_loop n# brings into the equational
context the equivalence \verb#add n Zero ≡ n#, which exactly corresponds to
the goal of the proof.

\begin{remark}
  The example \verb#add_n_Zero_loop# must be rejected because the underlying
  program (and hence proof structure) is non terminating (and hence not
  well-founded). The management of equivalences being correct by construction,
  incorrect equations can only be proved in a contradictory equational
  context. In the example, a faulty equation is learned when the
  non-well-founded recursive call is made.
\end{remark}

\subsection{Sized types}

The termination checking technology used by \pml is based on a notion of
size that is attached to typing judgments. In fact, inductive and coinductive
types are annotated using an ordinal size indicating the number of times
their definition can be unfolded (this is usual in the context of sized
types \cite{abel,hugues,sacchini13,subml}). Inductive or coinductive types,
such as lists or streams, can then be seen as sized types annotated by a large
enough (limit) ordinal.\footnote{In practice, $\omega$ is sufficient for most
of the usual data types, but this is not true in general.  Nonetheless, there
exists an ordinal that is large enough for all the inductive and coinductive
types to converge \cite{subml}.}

In practice, the implementation of \pml introduces sizes automatically when
typing recursive functions. This means that it replaces (some) inductive and
coinductive types, which carry a limit ordinal, with universal quantification
over all possible ordinals. Note that it is only possible to do so when the
obtained type is more general that the one that was given by the user. To
enforce this invariant, we only introduce quantification on inductive types
in negative position, and on coinductive types in positive position. In
practice, this heuristic works well on simple functions, but the user is
sometimes required to annotate the functions with explicit quantifications
for termination to be established. Moreover, manual annotation may lead to
more precise types, leading to more examples passing the termination check.
For example, the following \verb#map# function is accepted by the
implementation.
\begin{lstlisting}
val rec map : ∀a b, (a ⇒ b) ⇒ list⟨a⟩ ⇒ list⟨b⟩ =
  fun fn l {
    case l {
      Nil    → Nil
      Cns[c] → Cns[{hd = fn c.hd; tl = map fn c.tl}]
    }
  }
\end{lstlisting}
However, if the user writes a complex recursive function containing a
recursive call through the \verb#map# function, it will not be possible to
establish its termination (despite the fact that \verb#map# does not change
the size of the list it is applied to). To solve this problem, the user may
rather use a more precise sized type, which is a subtype of the former type.
\begin{lstlisting}
type slist⟨s:κ, a:ο⟩ = μ_s slist, [Nil ; Cns of {hd : a; tl : slist}]

val rec map : ∀s, ∀a b, (a ⇒ b) ⇒ slist⟨s,a⟩ ⇒ slist⟨s,b⟩ =
  fun fn l {
    case l {
      Nil    → Nil
      Cns[c] → Cns[{hd = fn c.hd; tl = map fn c.tl}]
    }
  }
\end{lstlisting}
\begin{remark}
  The type \verb~slist⟨s,a⟩~ should not be confused with the type of vectors
  \verb~vec⟨a,s⟩~ defined in Section~\ref{sect:mix}. Although they are both
  subtypes of regular lists, the former carries an ordinal $s$ (of sort
  $\kappa$) that can be used by the termination checker to establish size
  relations, while the latter contains a term (of sort $\tau$) corresponding
  to the size of the list (as computed by the \verb~length~ function) which
  cannot be used by the termination checker. Note however that these two types
  could be easily combined.
\end{remark}
\begin{remark}
  The above type of \verb#map# only enforces that the output list is at most
  as long as the input list. For instance, we could give the same type to a
  function taking the same two arguments and always returning an empty list.
\end{remark}
A similar scheme can be applied to \emph{insertion sort} for example, but not
for \emph{quick sort}, as discussed in previous work \cite{subml}. Indeed, a
richer language of ordinals would be required to express the fact that the
partition function preserves the number of elements of its input.\footnote{It
is nonetheless possible to show that quick sort is size-preserving using a
\pml proof.}

\subsection{Proof by equivalence to a terminating function}\label{sect:mc91}

Although the current implementation of \pml checks the termination of all
programs (not only proofs), it is possible to use the specific features of
the system to write termination proofs. Indeed, the equivalence relation
on which the system relies can be used to substitute one term with another,
provided that they are equivalent. This means that if we want to establish
the termination of a program that does not pass the termination check
directly, then we can instead establish the termination of any equivalent
program. We will here consider the example of the well-known \emph{McCarthy
91} function, whose termination cannot be established by most of the existing
termination criteria (if not all).
\begin{lstlisting}
include lib.nat
include lib.nat_proofs

def mccarthy91_hard =
  fix fun mccarthy91 n {
    if gt n u100 {
      minus n u10
    } else {
      mccarthy91 (mccarthy91 (add n u11))
    }
  }

// val mccarthy91 : nat ⇒ nat =
//   mccarthy91_hard
\end{lstlisting}
Note that here, the value \verb#mccarthy91_hard# is not defined as a usual,
type checked and termination checked value, but as a value object (using the
\verb#def# keyword). This means that this function can be manipulated as an
object of the logic, but not evaluated directly. Note also that we rely on
some functions (and constants) defined in the standard library of \pml. The
\verb#minus# function computes the difference, and the \verb#gt# function
tests whether its first argument is strictly greater than its second argument.

Although \pml is not able to prove the termination of the commented version
of \verb#mccarthy91#, we can give the following alternative (but equivalent)
definition.
\begin{lstlisting}
val mccarthy91_easy : nat ⇒ nat =
  fun n {
    if gt n u100 {
      minus n u10
    } else {
      u91
    }
  }
\end{lstlisting}
This second definition passes our termination check (it is not even
recursive), but it does not really correspond to the traditional definition of
the \emph{McCarthy 91 function}, which is a shame. We can nonetheless write a
\pml proof showing that these definitions are (pointwise) equivalent, which
will then allow us to replace one with the other. To do so, we first need to
show that \verb#mccarthy91_hard n# has value \verb#u91# for all numbers that
are not greater than \verb#u100#.
\begin{lstlisting}
val hard_aux: ∀n∈nat, gt n u100 ≡ false ⇒ mccarthy91_hard n ≡ u91 =
  fun n eq {
    {- ... -} // Can be done by enumerating the domain.
  }
\end{lstlisting}
We do not give the full proof for lack of space, but it can be easily
completed since the domain of quantification is finite. One simply needs
to explore the domain by pattern matching on \verb#n#, obtaining a trivial
proof for all numbers less or equal to \verb#u100#. In the case of numbers
greater that \verb#u100#, the presence of an additional successor produces
a contradiction with the hypothesis \verb#gt n u100 ≡ false# which allows
the enumeration to remain finite.
\begin{remark}
  This brute-force approach, although it could be easily automated, yields
  a proof that it rather inefficient. A better solution would be to write
  a proof by “induction”, which is what one would do on paper.
\end{remark}

Using the \verb#hard_aux# lemma, we can then show that the two implementations
of the \emph{McCarthy 91 function} produce the same result on every natural
number as follows.
\begin{lstlisting}
val hard_is_easy : ∀n∈nat, mccarthy91_easy n ≡ mccarthy91_hard n =
  fun n {
    use gt_total n u100;
    if gt n u100 {
      deduce mccarthy91_easy n ≡ minus n u10;
      deduce mccarthy91_hard n ≡ minus n u10;
      qed
    } else {
      deduce mccarthy91_easy n ≡ u91;
      show mccarthy91_hard n ≡ u91 using hard_aux n {};
      qed
    }
  }
\end{lstlisting}
The proof is straightforward\footnote{None of the \verb~deduce~ annotations
are necessary, they are only provided for clarity. The
\verb~gt\_total~ lemma is defined in the standard library, more detail about
its purpose will be given in the next section.} since the two implementations
have the same structure, and they share the same ``then'' branch. In the case
of the ``else'' branch, \verb#hard_aux# can be used to conclude. We can
then type check and prove the termination of the original version of the
\emph{McCarthy 91 function} as follows.
\begin{lstlisting}
val mccarthy91 : nat ⇒ nat =
  fun n {
    check mccarthy91_easy n  // Term used for type-checking.
      for mccarthy91_hard n  // Actual term used in the definition.
      because hard_is_easy n // Proof that they are equal.
    // The above really is "mccarthy91_hard n" (up to erasure).
  }
\end{lstlisting}
The annotation used in the definition of \verb#mccarthy91# instructs the
type-checker to substitute \verb#mccarthy_hard n# with \verb#mccarthy_easy n#
in the construction of the typing proof. This is only possible because these
two terms are equivalent (when \verb#n# has type \verb#nat#), as witnessed by
\verb#hard_is_easy n#. However, the term used for the computation will indeed
be \verb#mccarthy_hard n# after the annotations are erased.

\begin{remark}
  As all the types of \pml are closed under equivalence, it is always
  possible to replace a term by another equivalent term. This technique can
  not only be used for proving termination of functions such as
  \verb#mccarthy91#, but also for typing terms that would not be typable
  otherwise (but that are, for example, more efficient).
\end{remark}
\begin{remark}
  Note that we did not prove \verb#mccarthy_hard ≡ mccarthy_easy#, which may
  not even be true. Indeed, equivalence considers these two terms as untyped,
  and it is very well possible that they can be distinguished by a certain
  evaluation context.\footnote{In \pml, the equivalence \verb~t ≡ u~ being
  provable implies that \verb~t~ and \verb~u~ are observationally equivalent,
  which means that they have the same “observable behaviour” in every possible
  evaluation context. Note that we only observe termination, versus divergence
  or runtime error \cite{lepigrePhD,lepigre2016}.}
  A simpler example arises when comparing different implementations of the
  identity function on natural numbers: we have
  \verb#(fun n { case n { Zero ⇒ n | S[_] ⇒ n } }) k ≡ (fun n { n }) k# for
  all \verb#k# in \verb#nat#, but
  \verb#fun n { case n { Zero ⇒ n | S[_] ⇒ n } } ≡ fun n { n }# is false.
  These two functions can be distinguished using the argument \verb#false#,
  which yields a pattern matching failure on the former, while the latter
  successfully returns \verb#false#.
\end{remark}

\subsection{Internal totality proofs}

In this last section, we will give more explanations about the so-called
``totality proofs'' that are currently required in \pml. A function is said
to be \emph{total} if it computes some value, when applied to any value of
its domain. In \pml, the totality of functions can be expressed inside the
system using an existential quantification. We can thus write internal
totality proofs such as the following.
\begin{lstlisting}
val rec add_total : ∀n m∈nat, ∃v:ι, add n m ≡ v =
  fun n m {
    case n {
      Zero → qed
      S[k] → use add_total k m; qed
    }
  }
\end{lstlisting}
\begin{remark}
  Note that the value that is obtained by applying the function is not
  relevant here, nor is its type. We could however modify the definition
  of \verb#add_total# to make sure that an element of type \verb#nat# is
  returned. In this case, we could even use \verb#add_total# as \verb#add#.
\end{remark}

The reason why totality proofs are required in the system is strongly
related to the call-by-value evaluation strategy of the language. Indeed,
in call-by-value, a function can only be applied when all of its arguments
are values. More precisely, it only makes sense to reduce a $\beta$-redex
if the term in argument position is a syntactic value. To understand where
the notion of totality is really required, let us consider the following
proof example showing the associativity of addition.
\begin{lstlisting}
val rec add_assoc : ∀m n p∈nat, add m (add n p) ≡ add (add m n) p =
  fun m n p {
    use add_total n p;
    case m {
      Zero → qed
      S[k] → use add_assoc k n p; use add_total k n; qed
    }
  }
\end{lstlisting}
Ignoring the first call to \verb#add_total#, the proof starts by a case
analysis on variable \verb#m#. Let us consider the \verb#Zero# case, which
already illustrates very well the necessity for the totality proof. In this
branch, the automatic decision procedure learns the equation \verb#m ≡ Zero#.
As a consequence, the goal simplifies to
\verb#add Zero (add n p) ≡ add (add Zero n) p#, and even as
\verb#add Zero (add n p) ≡ add n p# since both \verb#Zero# and \verb#n#
are values (the function can thus be applied). However, the left-hand side
of the equation cannot reduce further because \verb#add n p# is not a value.
We can then only proceed using the totality proof produced by
\verb#add_total n p#, which gives us a value \verb#v# such that
\verb#add n p ≡ v#. As a consequence, this allows us to obtain
\verb#add Zero (add n p) ≡ add n p# as follows.
\begin{center}
  \verb#add Zero (add n p)  ≡  add Zero v  ≡  v  ≡  add n p#
\end{center}

\begin{remark}
  It is clear that the totality proof corresponding to a given function has
  a similar structure as the definition of the function itself. We may thus
  hope that totality proofs can be generated and called automatically, at
  least in most cases.
\end{remark}

\section{Future work}\label{future} 

The current implementation of \pml already allows for several convincing
examples, some of which cannot be expressed in other systems. They include,
for example, the \verb#extract# function of Section~\ref{classical} or the
\verb#mccarthy91# function of Section~\ref{sect:mc91}.
However, theoretical work and implementation work remain to be done for the
language to become fully practical, both as a programming language and as a
proof assistant.

\subsection{Mixing termination and non-termination}

As mentioned earlier, termination checking is only necessary
for \pml programs that are considered as proofs. In the theory, proving that
a program terminates amounts to showing that its typing derivation has a
well-founded circular structure. In this case, a standard semantic proof can
be used to prove normalisation \cite{subml}, the essential point being
that the adequacy of the type system can be established by induction on the
circular structure of proofs\footnote{Circularity is introduced by the typing
rule for the fixed-point combinator.}, provided that they are well-founded.

Alternatively, it is possible to type programs with standard (non-circular)
typing proofs,\footnote{This feature is not available in the current
implementation, which only accepts terminating programs.} to the expense of
losing normalisation since our termination criterion works by analysing the
circular structure of proofs. Note that lack of termination checking implies
the loss of soundness, but type-safety is nonetheless preserved.
As it is hard to automatically prove the termination of programs,
it is clear that a user will not want to be restricted to programs that can be
proved terminating. For this purpose, it is important to allow arbitrary
(type-safe) programs to be written, if only to prove them terminating later
(examples of such programs can be found in previous work \cite{pml}).

As programs that can be proven terminating can be typed in both ways, it is
natural to consider a way of mixing the two approaches in the theory. This
has actually already been implemented in a particular branch of our
implementation (called \emph{totality}) \cite{implem}. The corresponding
extension of the theory has also been checked informally.

\subsection{Other forms of effects, mutation}

One of the distinguishing features of \pml is the possibility for programs
to manipulate their own continuation (or evaluation context). This is
achieved using a construct similar to Scheme's \emph{call/cc}, or rather
Michel Parigot's $\mu$-abstraction \cite{parigot}, which triggers a form
of effect. As shown in Section~\ref{intro_classical}, it can be used to
realize theorems which only hold classically, by extracting a program from
their proofs \cite{raffalli_dickson}.

Although \pml is the first proof system based on a programming language with
effects and a classical realizability model \cite{lepigrePhD}, one may argue
that control structures only have a limited interest for writing practical ML
programs.\footnote{They can however be used to encode a form of exception
mechanism.} Other forms of effects however, for example input/output
directives or mutable cells, are essential to ML programmers. Although
it should be relatively easy to extend the system with the former, the latter
poses a real technical challenge. Indeed, it is not yet known how to account
for mutation in a classical realizability model.

\subsection{Subject reduction and strong safety}

The theory of \pml is based on a realizability model, which has the major
advantage of being flexible. More precisely, the adequacy lemma, which is
the keystone of the development, only needs to be modified locally to
encompass a new typing or subtyping rule. However, we have not yet
proved any subject reduction result for the system, and thus we only have a
weak form of type safety.

\subsection{Extensible variants and records (better inference)}

The current type-system of \pml requires a relatively small amount
of type annotations (at least for programs). Nonetheless, the system relies
on unification in several places, and it may happen that the system guesses
the wrong types. This situation arises most often with variant and record
types, for which some fields or constructors might be left out. This problem
can be solved using extensible variant types and record types, but we will
need to make sure that this does not pose any problem in the theory.

\subsection{Support for mutually recursive function}

In the current implementation, \pml lacks the possibility of defining
mutually recursive functions. Although it is always possible to encode
mutual recursion using additional parameters, this method does not perform
very well when combined with our termination checking technology. We
thus need to consider a different fixed-point instruction for our abstract
machine, which does not seem to pose any theoretical problem. The idea is
to replace the current fixed-point instruction with a term constructor
$\varphi a.v$, binding the term variable $a$ into the value $v$, and with the
reduction rule $\varphi a.v \to v[a := \varphi a.v]$.\footnote{Term variables
should not be confused with value variables (or $\lambda$-variables). In
particular, the former can be substituted with any term, while the latter can
only be substituted with values.} Mutual recursion can then be encoded using a
value $v$ that is a record containing several $\lambda$-abstractions, which
can still be typed using a simple unfolding rule (as in previous work
\cite{subml,lepigrePhD}).

\subsection{Certificates using proof traces for equivalences}

For now, it is not possible to formally check the proofs produced by \pml in
another system. Although the system already records the
proof trees that are produced during type-checking, the decision procedure
for program equivalence yet lacks the ability of producing a proof trace.
However, there is no theoretical evidence that it would not be possible for
the decision procedure to record enough information for an external prover
(for example Coq \cite{coq} or Dedukti \cite{dedukti}) to check the proofs
produced by \pml.

\section{Similar systems} 

To conclude this paper, we will compare \pml to other proof systems and
languages that can be used to formalise and prove program properties, or
that rely on similar principles.

\subsection{Dependent types in ML}

To our knowledge, the combination of call-by-value evaluation, side-effects
and dependent products has never been achieved before. At least not for a
dependent product fully compatible with effects and call-by-value. For
example, the Aura language \cite{aura} forbids dependency on terms that
are not values in dependent applications. Similarly, the $F^\star$ language
\cite{fstar} relies on (partial) let-normal forms to enforce values
in argument position. Daniel Licata and Robert Harper have defined a notion
of positively dependent types \cite{licata} which only allow dependency
over strictly positive types. Finally, in languages like ATS and DML
\cite{ats, dml}, dependencies are limited to a specific index language.

\subsection{Tools based on intuitionistic type theory}

The most actively developed proof assistants following the Curry-Howard
correspondence are Agda and Coq \cite{agda, coq}. The former is based on
Martin-Löf's dependent type theory and the latter on Coquand and Huet's
calculus of constructions \cite{coc, mltt}. These two constructive theories
provide dependent types, which allow the definition of very expressive
specifications. Contrary to \pml, Coq and Agda do not directly give a
computational interpretation to classical logic. Classical reasoning can
only be done through a negative translation or with the definition of axioms
such as the law of the excluded middle. In particular, these two languages
are not effectful. However, they are logically consistent, which means that
they only accept terminating programs. As termination checking is a difficult
(and undecidable) problem, many terminating programs are rejected. Although
this is not a problem for formalizing mathematics, this makes programming
tedious. In \pml, only proofs really need to be shown terminating, and it is
in any case possible to reason about non-terminating and even untyped
programs as they can be manipulated as objects in types.

\subsection{NuPrl and refinement types}

The NuPrl system \cite{nuprl} has many similarities with \pml on the
theoretical side, although it is inconsistent with classical logic. NuPrl
accommodates an observational equivalence relation
similar to ours (Howe's \emph{squiggle} relation \cite{howe}), which is
partially reflected in the syntax of the system. Being based on a Kleene-style
realizability model, NuPrl can also be used to reason about untyped terms.
Another major difference between \pml and NuPrl is that the latter is based on
refinement types, which means that it does not have an automatic way of
building typing derivations for programs. Indeed, typing derivations are built
interactively using a specific interface, and the user must say what typing
rule should be applied first.

\subsection{Partially consistent languages}

The TRELLYS project \cite{trellys} aims at providing a language in
which a consistent core interacts with type-safe dependently typed
programming with general recursion. Although the language is
call-by-value and effectful, it suffers from
value restriction like Aura \cite{aura}. The value restriction does not
appear explicitly but is encoded into a well-formedness judgement appearing
as the premise of the typing rule for application. Apart from value
restriction, the main difference between the language of the TRELLYS project
and ours resides in the calculus itself. Their calculus is Church-style (or
explicitly typed) while ours is Curry-style (or implicitly typed). In
particular, their terms and types are defined simultaneously, while our type
system is constructed on top of an untyped calculus.

\subsection{Systems aimed at program verification in ML}

Several systems have been proposed for verifying ML programs. ProPre
\cite{propre} relies on a notion of \emph{algorithms}, corresponding
to equational specifications of programs. It is used in conjunction with a
type system based on intuitionistic logic. Although it is possible to use
classical logic to prove that a program meets its specification, the
underlying programming language is not effectful.  Similarly, the PAF! system
\cite{baro} implements a logic supporting proofs of programs, but it is
restricted to a purely functional subset of ML. Another approach for reasoning
about purely functional ML programs is given in the work of Yann Regis-Gianas
\cite{regisgianas}, where Hoare logic is used to specify program properties.
Finally, it is also possible to reason about ML programs (including effectful
ones) by compiling them down to higher-order formulas
\cite{chargueraud10,chargueraud11}, which can then be manipulated using an
external prover such as Coq \cite{coq}.  In this case, the user is required to
master at least two languages, contrary to our system in which programming and
proving take place in a uniform framework.

\newpage
\bibliography{biblio}

\end{document}